\documentclass[useAMS,usenatbib]{mn2e}
\usepackage{amsmath}
\usepackage{amssymb}
\usepackage{graphicx}
\usepackage{subfig}
\usepackage{epsfig}
\usepackage{caption}
\usepackage{breqn}

\def\apj{ApJ}%
\title[Hunting Black Holes with Gaia]{Hunting Black Holes with Gaia}
\author[N. Mashian et al.]{Natalie Mashian$^{1}$\thanks{nmashian@g.harvard.edu}, Abraham Loeb$^{1}$\thanks{aloeb@cfa.harvard.edu} \\
$^{1}$Harvard-Smithsonian Center for Astrophysics, 60 Garden Street, Cambridge, MA 02138, USA}

\begin{document}

\pagerange{\pageref{firstpage}--\pageref{lastpage}} \pubyear{2017}

\maketitle

\label{firstpage}

\begin{abstract}
We predict the number of black holes with stellar companions that are potentially detectable with Gaia astrometry over the course of its five-year mission. Our model estimates that nearly 2$\times$10$^5$ astrometric binaries hosting black holes and stellar companions brighter than Gaia's detection threshold, $G \sim$ 20, should be discovered with 5$\sigma$ sensitivity. Among these detectable binaries, systems with longer orbital periods are favored, and black hole and stellar companion masses in the range $M_{BH} \sim$ 6 - 10 M$_\odot$ and M$_* \sim$ 1 - 2 M$_\odot$, respectively, are expected to dominate.
\end{abstract}



\section{Introduction}
Stellar evolution models, chemical enrichment by supernovae within the Milky Way, and gravitational microlensing events all indicate that a population of about $\sim$10$^8$-10$^9$ stellar-mass black holes resides in our Galaxy \citep{1983bhwd.book.....S,vanden,1994ApJ...423..659B,1998ApJ...496..155S,2002ApJ...576L.131A}. And yet, despite such high estimates, fewer than fifty stellar black hole candidates have been studied and confirmed, all of which are in X-ray binary systems \citep{2006csxs.book.....L,2007IAUS..238....3C,2013MNRAS.430.1538F,2013arXiv1312.6698N}. We are therefore left to wonder, where are all these ``missing" black holes and how can we observe them?

While black holes in binary systems with ongoing mass transfer have been preferentially detected via X-rays, we expect that at any given moment, a dominant fraction of stellar black holes reside in binaries with no substantial X-ray emission. The binary system may be an X-ray transient, undergoing dramatic episodes of enhanced mass-transfer followed by quiescent phases where the X-rays switch off and the flux drops by several magnitudes; all known low mass black hole binaries (LMBHBs) fall into this category, demonstrating transient behavior \citep{1995xrbi.nasa..126T,1995xrbi.nasa....1W,mcclintock}. \citet{1999ApJ...513..811M} suggested that there may be a population of persistent LMBHBs that are simply dim due to the very low radiative efficiencies of their advection-dominated accretion flow zones. These systems lack large-amplitude outbursts and thus remain difficult to detect with their persistent, yet faint X-ray emission.

There may be a significant number of stellar black holes residing in \textit{detached} binary systems with companion stars that have not yet reached, or may never reach, the evolutionary stage of transferring mass by Roche-lobe overflow to their compact primaries \citep{2006A&A...454..559Y}. \citet{2001AstL...27..645K} suggests the existence of hundreds of times more detached binaries composed of a massive OB star and a stellar-wind accreting black hole that remain impossible to detect through X-ray observations. The spherically symmetric accretion that takes place in these systems can only result in the effective generation of a hard radiation component if equipartition is established between the gravitational and magnetic energies in the flow. Given the magnetic exhaust effect operating in these binaries, such equilibrium is rarely established and most black holes in these systems remain unobservable \citep{2001AstL...27..645K}.

Given the difficulties associated with detecting stellar-mass black holes in binary systems with no substantial X-ray emission, we explore the possibility of discovering these elusive, compact objects through astrometric observations. 
\citet{2016MNRAS.458.3012W} demonstrate this ability to detect dark stellar remnants through parallax microlensing events, particularly with the aid of future Gaia microlensing astrometric measurements.
The recent advent of the European Space Agency mission Gaia \citep{2001A&A...369..339P} is expected to transform the field of astrometry by measuring the three-dimensional spatial and velocity distribution of nearly $\sim$ 1 billion stars brighter than magnitude $G \sim$ 20 \citep{2016A&A...595A...4L}. Over the course of its five-year mission lifetime, Gaia is expected to perform an all-sky survey, observing each source an average of 70 times and yielding final astrometric accuracies of roughly 10 $\mu$as (micro-arcsec) at $G \sim$ 13 mag, 30 $\mu$as at $G \sim$ 15, and 600 $\mu$as at $G \sim$ 20 \citep{2016A&A...595A...1G}. By surveying an unprecedented 1\% of the Galaxy's total stellar population with unparalleled precision, Gaia will not only confirm and improve upon observations of the few dozen active X-ray binaries that have already been closely studied, but it will also provide a unique opportunity to detect astrometric binaries that have one invisible component, i.e. a black hole, by observing the motion of the visible companion around the system's barycenter.


In this Letter, we seek to estimate the number of black holes with stellar companions that are potentially detectable with Gaia astrometry. We outline the components of our model and explore the various constraints on the detectability of these binary systems in section \S2. Our results are presented and discussed in section \S3. We adopt a flat, $\Lambda$CDM cosmology with $\Omega_m$ = 0.3, $\Omega_\Lambda$ = 0.7, and $H_0$ = 70 km\,s$^{-1}$Mpc$^{-1}$, consistent with the most recent measurements from the \textit{Planck} satellite \citep{2016A&A...594A..13P}.

\section{The Model}

To calculate the number of black holes we expect Gaia to detect astrometrically over its five-year lifetime, our model accounts for several limiting factors: (\textit{i}) the fraction of stars assumed to end their lives as a black hole in a binary system with a fellow stellar companion; (\textit{ii}) the fraction of those systems with stellar companions that are still shining and whose astrometric wobbles are thus detectable; (\textit{iii}) the fraction of those systems with periods in the range that Gaia can observe given its mission lifetime and astrometric precision; and (\textit{iv}) the volume of space Gaia can probe given its detection threshold of $G \sim$ 20. We will now briefly discuss each factor in turn.

\vspace{10pt}
\noindent(\textit{i}) \textit{Black holes with stellar companions}

\vspace{1pt}\noindent The massive star progenitors of stellar black holes are commonly found in binaries, with galactic surveys yielding bias-corrected spectroscopic binary fractions of 0.6-0.7 in the Milky Way \citep{2007ApJ...670..747K,2012Sci...337..444S,2015A&A...580A..93D,2017arXiv170301608S}. \citet{2012ApJ...751....4K} suggest that the binary fraction among these massive stars may even be as high as 90\% for systems with periods up to 10$^4$ years. The question then becomes whether these black hole progenitors keep their companions throughout the stellar evolutionary process. Various phenomena can lead to the unbinding of a binary, including dynamical interactions, mass loss, supernova explosions of either companion, and merging of the two sources. \citet{2013MNRAS.430.1538F} estimate that about one quarter of O-stars merge with their companions during evolution and that nearly 50\% of black holes have companions that undergo iron core-collapse supernovae (CCSNe), producing neutron stars with natal kicks that can unbind the system with their large velocities. 

The formation of the black hole itself may disturb the binary system depending on the formation mechanism.
For the lower mass end of black hole progenitors, a metastable proto-neutron star (PNS) is left behind by a weak CCSN and a black hole appears after fallback accretion of the part of the stellar envelope that was not successfully expelled by the supernova (SN) explosion. In the case of more massive stars with larger iron cores, successful SNe explosions are even more difficult to achieve and black holes are formed through accretion of the entire stellar material, either through direct collapse or as a result of failed SNe \citep{2008ApJ...684.1336K,2010ApJ...714.1217B,2016arXiv160901283A}. 

This notion that some massive stars undergo collapse without producing an explosion was first invoked by theorists to explain away difficulties in producing SN explosions in analytical models and hydrodynamic simulations \citep{1986ARA&A..24..205W,2002ApJ...572..944G}. Since then, failed SNe have gained ground as a possible mechanism for producing cosmological gamma-ray bursts \citep{1993ApJ...405..273W} and as a potential solution to the ``Red Supergiant Problem", referring to the statistically significant absence of higher mass ($\gtrsim$ 16 M$_\odot$) red supergiants exploding as SNe \citep{2009ARA&A..47...63S,2015PASA...32...16S}. Evidence in line with the suggestion that $\gtrsim$ 16 M$_\odot$ stars do not explode as optically bright SNe has been mounting in recent years. To date, there have been no observed Type IIP SNe with strong [O{\scriptsize II}] emission consistent with a zero-age main sequence (ZAMS) progenitor mass $\gtrsim$ 16 M$_\odot$ \citep{2014MNRAS.439.3694J}. Furthermore, \citet{2013ApJ...769...99B} found that Galactic abundances can be reproduced even if no stars with masses above 25 M$_\odot$ contribute metals via SN explosions, a mass threshold that can be even lower if uncertain mass loss parameters and reaction rates are modified. 
Though simulations suggest that there is no single mass below which all stars explode and above which black holes form by implosion, typically 95\% of stars that do explode have masses less than 20 M$_\odot$ \citep{2016ApJ...821...38S}. For those stars that end up as black holes, which are most but not all stars with masses above 20 M$_\odot$, only a few were found to form black holes via fallback \citep{2016ApJ...821...38S}. 

Given the theoretical predictions and supporting observations, we assume for the purposes of our calculation that progenitor stars with mass $\geq$ 20 M$_\odot$ will collapse to form black holes without undergoing an explosion.
The mass loss experienced by these black hole progenitors may effectively disrupt the binary if the system loses more than half of its mass suddenly during stellar collapse; in such cases, the newly formed black hole may receive a natal kick, enhancing binary disruption \citep{2012MNRAS.425.2799R,2013ApJ...769..109L,2015MNRAS.446.1213K}. On the other hand, if the progenitor loses its mass gradually, i.e. via wind, adiabatic invariants should be preserved and the binary separation will adjust accordingly to accommodate the mass loss, allowing the binary to remain gravitationally bound \citep{2014ARA&A..52..487S}. 
The final fraction of stellar objects assumed to end their lives as a black hole in a binary system is thus,
\begin{equation}
f_{B.F.}=f\int_{20 \textrm{ M}_\odot}^{150 \textrm{ M}_\odot}\hspace{-3pt}\epsilon(M_*)dM_* \,\,\,,
\end{equation}
where $f$ is fudge factor accounting for the fraction of black hole progenitor stars that either never host a companion, or lose their companion over the course of stellar evolution.
Rapid mass loss and SN-like kicks during black hole formation will result in small values of $f$, while formation via implosion or failed SN of stripped Wolf-Rayet stars imply values closer to unity.

\begin{figure*}
\includegraphics[width=470pt,height=190pt]{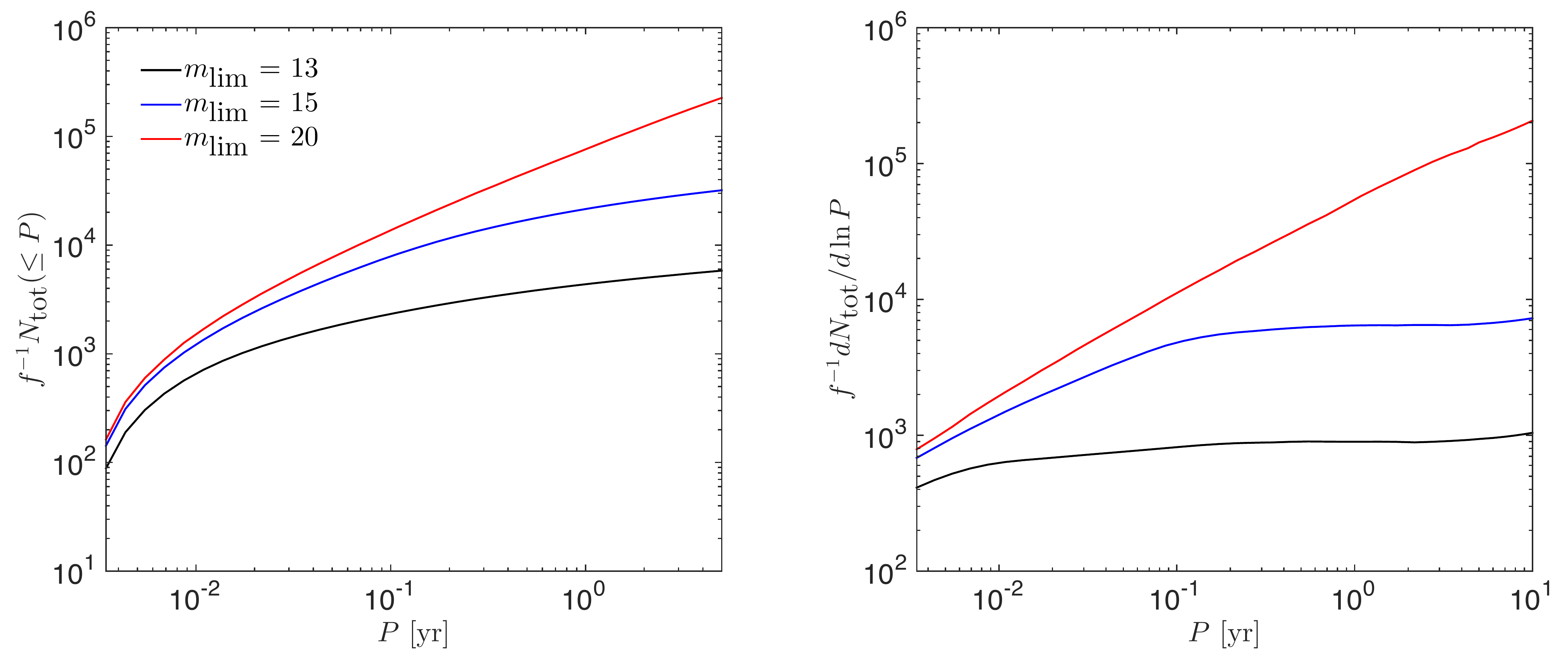}\\
\caption{\textit{Left panel}: Cumulative number of detectable astrometric binaries hosting black holes with orbital periods less than $P$, assuming detection thresholds of $m_{\textrm{lim}}$ = 13, 15, and 20 in the $G$ band. \textit{Right panel}: The corresponding differential count per logarithmic interval in $P$. Numbers are shown for a 5$\sigma$ detection and may be modified by a multiplicative factor $f$ to account for the fraction of black hole progenitor stars that end their lives isolated.}
\end{figure*}

The initial mass function (IMF) of progenitor stars, $\epsilon(M_*)$, is a multi-part power law taken from \citet{2007astro.ph..3124K},
\[
\epsilon(M_*)=k
\begin{cases} 
          \left(\frac{M_*}{0.5}\right)^{-1.3} & 0.1< M_*\leq 0.5 \textrm{ M}_\odot \\
      \left(\frac{M_*}{0.5}\right)^{-2.3} & 0.5 <M_*\leq 150 \textrm{ M}_\odot
   \end{cases}
\]
where $k$ is set to normalize the distribution such that $\int_{0.1 \textrm{ M}_\odot}^{150 \textrm{ M}_\odot}dM_*\epsilon(M_*)=1$.

\vspace{20pt}
\noindent(\textit{ii}) \textit{Visible binary companions}

\vspace{2pt}\noindent In order to detect a black hole via the astrometric wobble of its stellar companion, the star must still be shining, i.e. observable in the visual range. To determine the fraction of binary systems with stellar sources that are currently burning fuel, we adopt the simplifying assumption that the Milky Way stars follow the cumulative star-formation history of the Universe,
\begin{equation}
\rho_*(z)=(1-R)\int_0^t\dot{\rho_*}(t)dt=(1-R)\int_z^\infty\frac{\dot{\rho_*}(z')}{H(z')(1+z')}dz' \,\,\,,
\end{equation}
where $R$ = 0.39 is the ``return fraction", $H(z')=H_0\left[\Omega_m(1+z')^3+\Omega_\Lambda\right]^{1/2}$, and the best-fitting comoving star formation rate density \citep{2016arXiv160607887M} is
\begin{equation}
\dot{\rho_*}(z)=0.01\frac{(1+z)^{2.6}}{1+\left[(1+z)/3.2\right]^{6.2}} \,\, \textrm{M$_\odot$\,yr$^{-1}$\,Mpc$^{-3}$} \,\,\,.
\end{equation}
If we express the fraction of baryons locked up in stars at a certain redshift as $f_*(z)=\rho_*(z)/\rho_b$ where $\rho_b=\Omega_b\rho_{crit}$, then the fraction of stars with ZAMS mass $M_*$ that are still ``alive" today is given by
\begin{equation}
f_{\textrm{shining}}(M_*)=1-\frac{\rho_*(z(t_{LB}=t_{\textrm{age}}(M_*)))}{\rho_*(0)}
\end{equation}
where the look-back time, $t_{LB}$, is set to the stellar lifetime, $t_{\textrm{age}}(M_*)$. Stellar lifetimes were calculated using MIST models and the MESA stellar evolution libraries \citep{2011ApJS..192....3P,2016ApJ...823..102C,2016ApJS..222....8D} and $f_{\textrm{shining}}(M_*)$ is set to unity for stars with lifetimes that exceed the age of the Universe.

\vspace{10pt}
\noindent(\textit{iii}) \textit{Accessible range of periods}

\vspace{2pt}\noindent Binary surveys that probe massive stars often find a uniform distribution in log period, following what is commonly referred to as \"Opik's law \citep{opik,1980ApJ...242.1063G,2007A&A...474...77K}. We therefore adopt a log-flat distribution, $p(\log P) \propto (\log P)^\gamma$ with $\gamma$ = 0, to characterize the orbital period distribution of the binary systems under consideration. 

For the astrometric signals of these binaries to be detectable by Gaia, the orbital periods should not exceed Gaia's mission lifetime, $P_{\textrm{max}}$ = $t_G$ = 5 years. There is also a lower bound on the range of accessible periods that is set by 
Gaia's astrometric precision at a given magnitude, $\alpha_{\textrm{G}}(m)$ \citep{2016A&A...595A...1G}. The astrometric signature of the binary, which must be at least as large as $\alpha_{\textrm{G}}$, is given by
\begin{equation}
\alpha=\frac{a_*}{r}=\frac{M_{BH}}{M_{BH}+M_*}\frac{a}{r}
\end{equation}
where $r$ denotes the distance of the binary system from the Sun, $a_*$ is the observable semi-major axis of the star's orbit projected onto the sky, and $a$ is the corresponding semi-major axis of the relative orbit of the black hole and companion star, with masses $M_{BH}$ and $M_*$ respectively. Since $a$ is related to the period $P$ through Kepler's third law, the constraint on the astrometric signature, $\alpha \geq \alpha_{G}$, translates into a minimum detectable orbital period of
\begin{equation}
P_{\textrm{min}}=(M_{BH}+M_*)\left(\frac{r\,\alpha_{\textrm{G}}(m)}{M_{BH}}\right)^{3/2} \,\,\,.
\end{equation}
The Gaia data release has thus far only provided proper motions and parallaxes derived from the global astrometric solution. Since the effects of parallax are similar to those of a stellar wobble in a binary orbit, we adopt the parallax standard error (eqs. (4)-(6) in \citet{2016A&A...595A...1G}) as a proxy for $\alpha_G$, the astrometric accuracy; future data treatment will allow us to compute the orbital elements of binaries and break any degeneracies between these two effects.

Given that massive binary systems with orbital periods less than 1 day are extremely rare and more likely to be disruptively interactive, we set the minimum orbital period equal to 1 day if $P_{\textrm{min}}$ in equation (6) drops below this threshold.

\begin{figure*}
\includegraphics[width=450pt,height=180pt]{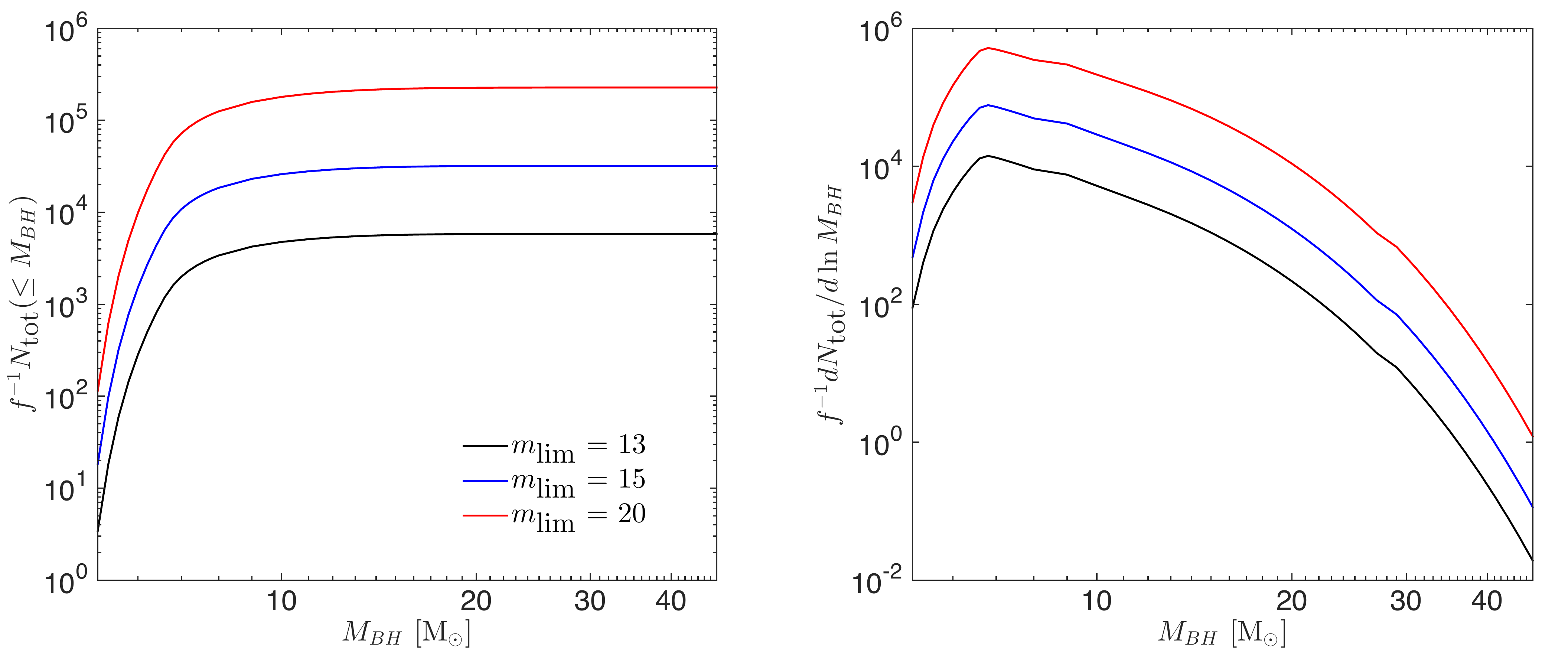}\\
\includegraphics[width=450pt,height=180pt]{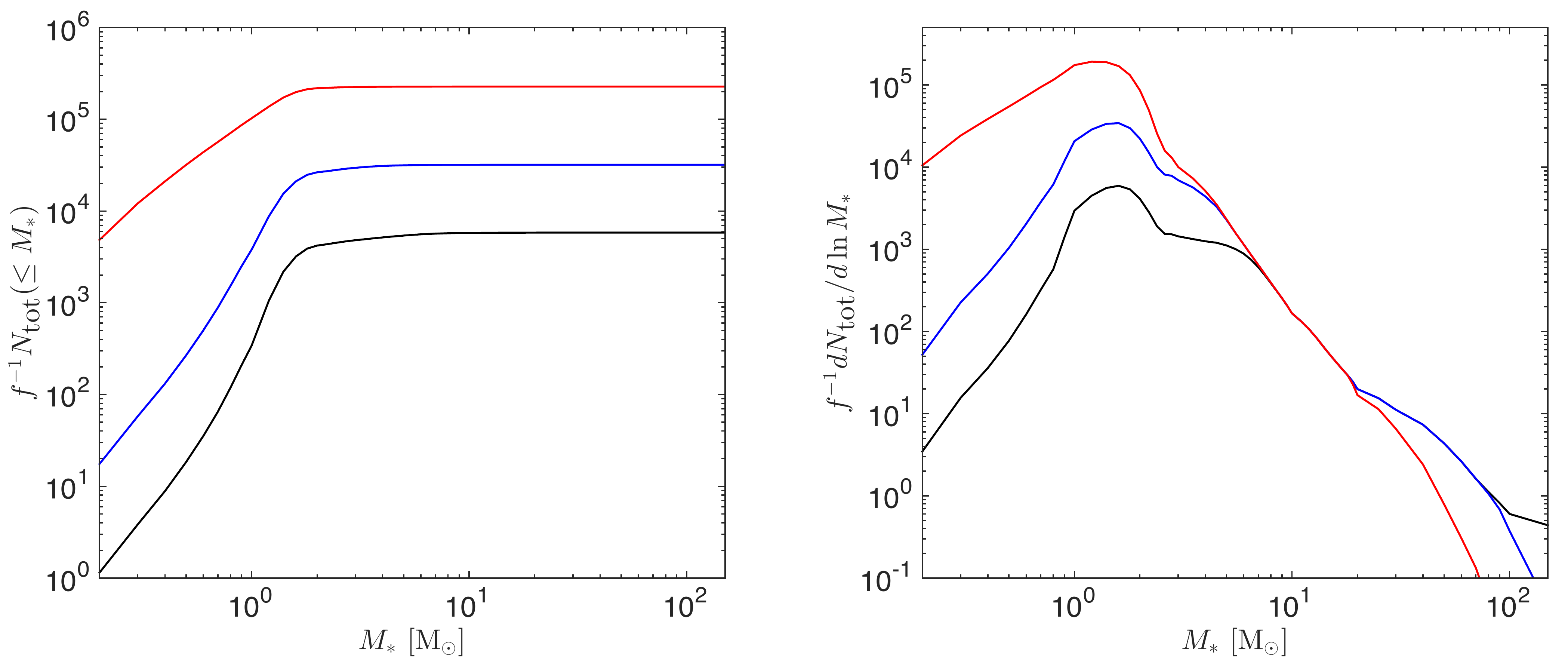}
\caption{\textit{Left panel}: Cumulative number of detectable astrometric binaries hosting black holes with mass less than $M_{BH}$ (\textit{top}) and stellar companions with mass less than $M_*$ (\textit{bottom}). \textit{Right panel}: The corresponding differential count per logarithmic interval in mass, assuming 5$\sigma$ detection. The multiplicative factor $f$ accounts for the fraction of isolated black hole progenitors with no stellar companion.}
\end{figure*}

\vspace{10pt}
\noindent(\textit{iv}) \textit{Volume of space probed by Gaia}

\vspace{2pt}\noindent The final factor our model accounts for is the fraction of stars Gaia can probe given its detection threshold, $m_{\textrm{lim}}$. A star of mass $M_*$ with a corresponding luminosity of $L(M_*)$ can be observed out to a distance
\begin{equation}
d_{\textrm{max}}(M_*)=0.3\sqrt{10^{(m_{\textrm{lim}}+2.72)/2.5}L(M_*)} \,\,\,\textrm{pc}
\end{equation}
where $L(M_*)$ is computed using the previously mentioned MIST models (in units of solar luminosity). We model the stellar density profile with a double exponential thin and thick disk centered at the solar position in the Galaxy, $(R_\odot, Z_\odot)$ = (8 kpc, 25 pc), with scale lengths of $h_{R,\textrm{thin}}$ = 2.6 kpc and $h_{R,\textrm{thick}}$ = 3.6 kpc, scale heights of $h_{z,\textrm{thin}}$ = 0.3 kpc and $h_{R,\textrm{thick}}$ = 0.9 kpc, and a normalization of the thick disk (relative to the thin) of $f_{\textrm{thick}}$ = 0.04 \citep{2008ApJ...673..864J,yoshii}. Normalized further to yield a total stellar mass of $M_{*,\textrm{tot}}$ = 6.08$\times$10$^{10}$ M$_\odot$ \citep{2015ApJ...806...96L}, this stellar number density distribution takes the following form (in spherical coordinates),
\begin{equation}
\begin{split}
n(r,\theta,\phi)&=n_0\left[\exp{\left(-\frac{(r\sin{\phi}+R_\odot)}{2.6 \textrm{ kpc}}-\frac{|r\cos{\phi}+Z_\odot|}{0.3 \textrm{ kpc}}\right)}\right.\\
&+\left.0.04\exp{\left(-\frac{(r\sin{\phi}+R_\odot)}{3.6 \textrm{ kpc}}-\frac{|r\cos{\phi}+Z_\odot|}{0.9 \textrm{ kpc}}\right)}\right]
\end{split}
\end{equation}
where $n_0 \approx$ 3 pc$^{-3}$. 

\vspace{10pt}
\noindent Taking into account all these various limiting factors, the resulting total number of black holes we can expect Gaia to observe astrometrically over the course of its five-year mission takes the following form,
\begin{equation}
\begin{split}
N_{\textrm{tot}}&=f_{B.F.}\int_{5 \textrm{ M}_\odot}^\infty \hspace{-10pt} dM_{BH} \,\psi(M_{BH})\,\,\,\times\\
&\int_{0.1 \textrm{ M}_\odot}^{150 \textrm{ M}_\odot}\hspace{-15pt} dM_* \,f_{\textrm{shining}}(M_*)\,\epsilon(M_*)\int_0^{2\pi}\hspace{-5pt}d\theta \int_0^\pi d\phi\,\sin{\phi}\,\,\,\times\\
&  \int_0^{d_{\textrm{max}}(M_*,m_{\textrm{lim}})}\hspace{-15pt}dr\,r^2 \int_{\log{\left(\textrm{Max}[1 \textrm{ day, } P_{\textrm{min}}(r,M_*,M_{BH})]\right)}}^{\log{\left(t_G\right)}}\hspace{-100pt} d\left(\log{P}\right)\,n(r,\theta,\phi)\,p\left(\log{P}\right)\,\,\,,
\end{split}
\end{equation}
where $\psi(M_{BH})$ represents the black hole mass distribution adopted from \citet{2010ApJ...725.1918O,2012ApJ...757...55O},
\begin{equation}
\psi(M_{BH})=\left\{A(M_{BH})^\delta+\left[B(M_{BH})^{-\delta}+C(M_{BH})^{-\delta}\right]^{-1}\right\}^{1/\delta} \,\,\,,
\end{equation}
with $\delta$ = -10.0 and,
\begin{eqnarray}
&A(M_{BH})&=4.367-1.7294M_{BH}+0.1713M^2_{BH}\nonumber\\
&B(M_{BH})&=14.24e^{-0.542M_{BH}}\nonumber\\
&C(M_{BH})&=3.322e^{-0.386M_{BH}} \,\,\,.
\end{eqnarray}

\section{Results \& Discussion}
Using the model outlined above, we estimate the number of astrometric binaries with stellar black holes that Gaia can potentially detect with 5$\sigma$ sensitivity. Figure 1 depicts the cumulative number of expected binaries with orbital periods less than some value $P$, assuming various limiting magnitudes in the $G$ band, $m_{\textrm{lim}}$. Out of the nearly 1 billion stellar sources  that will be observed by Gaia, we expect to find $\sim$ 5800 binaries with periods $\lesssim$ 5 years that host a black hole ($M_{BH} \geq$ 5 M$_\odot$) and a stellar companion brighter than $G \sim$ 13. The distribution of orbital periods among these potentially detectable binaries with $G \lesssim$ 13 will be roughly uniform across the range 4 days $\lesssim P \lesssim$ 5 years.
Alternatively, if we observe stars as faint as $G \sim$ 20, corresponding to the magnitude threshold of Gaia, this number rises to $N_{\textrm{tot}}(P \leq \textrm{ 5 yr})$ = 2$\times$10$^5$, 0.02\% of the surveyed sources. Correspondingly, inclusion of these fainter sources results in a period distribution among the detected astrometric binaries that is more heavily weighted towards longer orbital periods (Figure 1, right panel). 
If the mission lifetime is extended to 10 years, Gaia may observe up to 3.4$\times$10$^5$ astrometric binaries hosting stellar black holes.
We also note that all of these conservative estimates for a 5$\sigma$ detection may be modified by the fudge factor, $f$, introduced in \S2, that accounts for black holes that end up `isolated', either because their progenitor stars never hosted a companion or because they lost their companion over the course of stellar evolution.

Figure 2 illustrates how these numbers vary instead with the mass of the black hole (\textit{top panel}) and stellar companion (\textit{bottom}). The differential count per logarithmic interval in mass (plotted in the right panel), demonstrates that a significant fraction of the detected astrometric binaries are expected to have black hole masses between 6 and 10 M$_\odot$, reflecting the intrinsic black hole mass distribution which peaks in the range 5-7 M$_\odot$ and rapidly decreases at larger masses \citep{2010ApJ...725.1918O,2011ApJ...741..103F}. The stellar companions in these detectable astrometric binaries are expected to have masses predominately in the solar range, with $dN/d\ln{M_*}$ peaking around $\sim$ 1-2 M$_\odot$.

Astrometric observations provided by Gaia will yield the period of the orbital motion around a binary's barycenter, as well as $\alpha$, the angular semi-major axis of the luminous companion's orbit. Kepler's third law, expressed in terms of these quantities, takes on the form,
\begin{equation}
\left(\frac{P}{\textrm{1 yr}}\right)^{-2}\left(\frac{\alpha \,r}{\textrm{1 AU}}\right)^3=\frac{M_{\textrm{dark}}^3}{(M_{\textrm{dark}}+M_{*})^2 M_\odot} \,\,\,,
\end{equation}
where $M_{\textrm{dark}}$ and $M_{*}$ are the masses of the dark component and luminous stellar component, respectively. Astrometric and parallax measurements yield $P$, $\alpha$, and $r$, while Gaia's low-resolution spectrographs will yield spectral typing for most stars, allowing $M_*$ to be calculated in many cases and the mass of the unseen companion, $M_{\textrm{dark}}$, to be easily derived via eq. (12). Stellar-mass black holes can thus be identified astrometrically in systems where the proper motion of the star, implies the existence of a dark companion with $M_{\textrm{dark}} \gtrsim$ 3 M$_\odot$.

The invisible companions of astrometrically observed metal-poor, low-mass stars ([Fe/H] $<$ -1, $M_* <$ 1 M$_\odot$) may be stellar remnants from the dawn of the Universe, offering to shed light on the formation of the first stellar black holes in the early stages of galaxy assembly and evolution. Conversely, astrometric binaries with a stellar black hole and a fellow high-mass stellar companion ($M_* >$ 20 M$_\odot$) offer candidate systems for future gravitational-wave observations with the Laser Interferometer Gravitational Wave Observatory (LIGO) or eLISA \citep{2017arXiv170101736C}. By providing astrometric observations of nearly 1 billion galactic stars with unprecedented precision, Gaia promises to probe the stellar black hole population, a population that, despite its expected abundance, has mostly evaded detection thus far.

\section{Acknowledgements}
This work was supported by the Black Hole Initiative, which is funded by a grant from the John Templeton Foundation. We thank C. Kochanek and J. McClintock for helpful comments on the manuscript.

\label{lastpage}
\end{document}